# Prediction of Potential Commercially Available Inhibitors against SARS-CoV-2 by Multi-Task Deep Learning Model

Fan Hu, Jiaxin Jiang, Peng Yin*

Guangdong-Hong Kong-Macao Joint Laboratory of Human-Machine Intelligence-Synergy Systems, Shenzhen Institutes of Advanced Technology, Chinese Academy of Sciences, Shenzhen 518055, China

*To whom correspondence should be addressed

## Abstract

The outbreak of COVID-19 caused millions of deaths worldwide, and the number of total infections is still rising. It is necessary to identify some potentially effective drugs that can be used to prevent the development of severe symptoms or even death for those infected. Fortunately, many efforts have been made, and several effective drugs have been identified. The rapidly increasing amount of data is of great help for training an effective and specific deep learning model. In this study, we propose a multi-task deep learning model for the purpose of screening commercially available and effective inhibitors against SARS-CoV-2. First, we pretrained a model on several heterogenous protein–ligand interaction datasets. The model achieved competitive results on some benchmark datasets. Next, a coronavirus-specific dataset was collected and used to fine-tune the model. Then, the fine-tuned model was used to select commercially available drugs against SARS-CoV-2 protein targets. Overall, twenty compounds were listed as potential inhibitors. We further explored the model interpretability and observed the predicted important binding sites. Based on this prediction, molecular docking was also performed to visualize the binding modes of the selected inhibitors.

# Introduction

As of June 2022, SARS-CoV-2 has sickened more 5 hundred million and killed over 6 million people across the globe. SARS-CoV-2 is the seventh member of the family of coronaviruses that infect humans[1,2]. Similar to MERS-CoV and SARS-CoV, SARS-CoV-2 causes severe respiratory diseases and is capable of spreading from person to person. Unfortunately, the rapid mutation of this virus makes it difficult to develop effective vaccines, especially for Omicron. Nobody knows when the pandemic will finally end. Therefore, continuous development of effective anti-SARS-CoV-2 drugs is necessary to prevent the worsening of symptoms and even death of infected people. As a positive-sense, single-stranded RNA beta-coronavirus, SARS-CoV-2 encodes structural, non-structural and accessory proteins. Among these, RNA-dependent RNA polymerase (RdRp), 3-chymotrypsin-like protease (3CLpro), Papain-like protease (PLpro), helicase and spike glycoprotein are supposed to be the main targets. Several compounds that targeted these viral proteins and inhibited coronavirus in vitro have been reported and moved into clinical trials[3]. For example, remdesivir is an approved HIV reverse transcriptase inhibitor, which has broad-spectrum activities against RNA viruses such as MERS-CoV and SARS-CoV, but shows less effective in a Ebola clinical trial[4–6]. Previous reports showed that remdesivir inhibited SARS-CoV-2 in vitro with EC50=0.77μM[7], and was used to treat a SARS-CoV-2 infected patient in the United States[8]. Yet more potential inhibitors against SARS-CoV-2 are still needed.

Computational methods, such as molecular docking, select compounds that can bind to target protein and thus improve the success rates of drug discovery. Recently, methods based on deep learning have gained impressive performance on protein-ligand binding prediction[9–11]. One main advantage of this algorithm is that it can extract hidden features automatically from raw data and thus significantly improve the prediction accuracy. Yet deep learning model might suffer from the generalizability issue due to lack of data. Recently, pretraining model on a large-scale dataset before applying to a small dataset has emerged as a powerful paradigm for solving this issue[12].

In this study, we propose a multi-task deep learning model for selecting potential SARS-CoV-2 inhibitors. First, the model was pretrained on various heterogenous protein-ligand interactions datasets. The model has achieved competitive results on several protein-ligand benchmark datasets. Next, a coronavirus-specific dataset was collected and used to fine-tune the model. The fine-tuned model was then used to select available commercial drugs against several important proteins of SARS-CoV-2. Overall, twenty drugs were listed as potential inhibitors. Furthermore, we explored the model interpretability and found that the predicted important binding sites were close to ground truth.

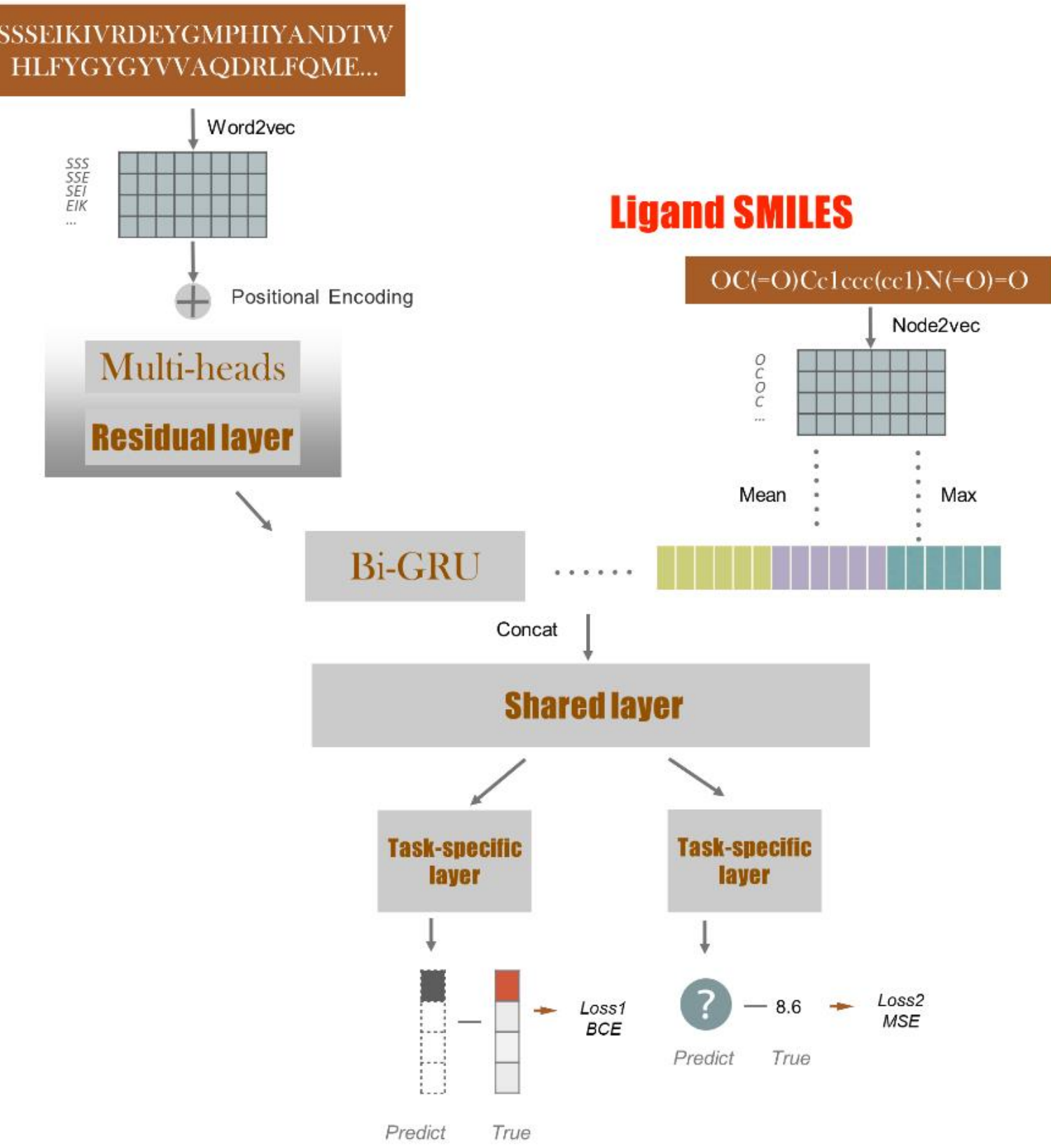

Fig. 1 Schematic of the proposed model. The model involves two parts: protein/ligand features extraction and their interaction prediction. First, protein sequence is processed in turn by word2vec[13], multi-heads residual layer and Bi-GRU modules. Ligand smiles is processed by node2vec[14]. Then, their representations are fed into shared layer and task-specific layer to output results according to the labels.

# Results and discussion

## Model performance on benchmarks

We first trained a multi-task deep learning model on various heterogenous protein-ligand interaction datasets. The model has achieved competitive results on several benchmark sets. For example, the PDBbind v.2016 dataset is a widely used benchmark for evaluating protein-ligand interaction prediction methods[15]. It provides more than ten thousand protein-ligand structural complexes with binding affinity (e.g., $K_d$, $K_i$). We split the PDBbind set the same way as Pafnucy[16], a classic deep learning-based method for predicting protein-ligand affinity. Moreover, two independent test sets CASF-2013[17] and Astex Diverse set[18] were used to test the generalizability of the model. Two metrics, RMSE (root mean square error) and Pearson's correlation coefficient R, were used to evaluate the regression task. As shown in Fig. 2, the model achieves RMSE=1.538, R=0.71 on the test set and performs well on both two independent tests. These results are better than most classic structure-based methods evaluated on the same set. It is worth noting that our model is only sequence-based.

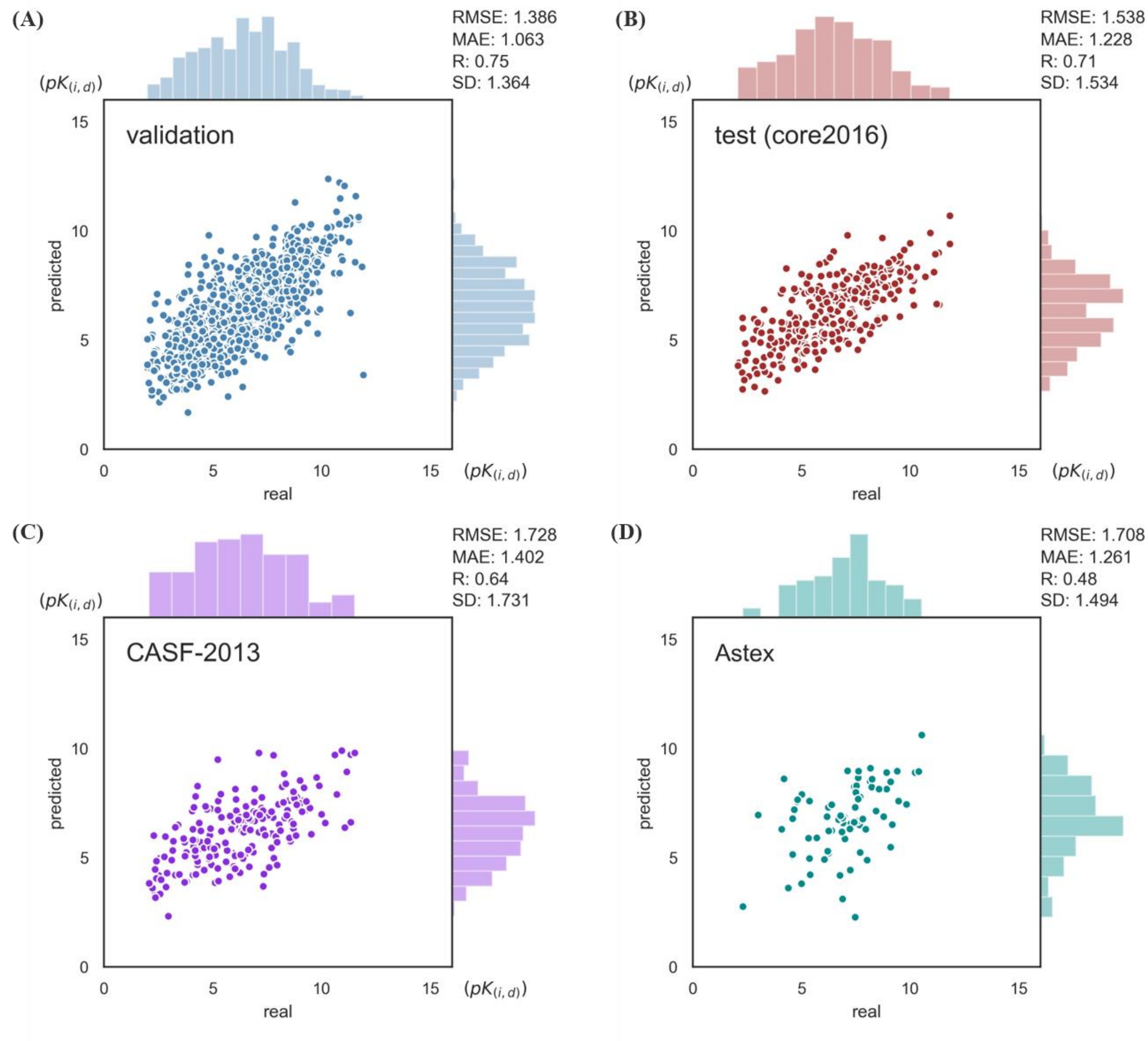

Fig. 2 Model performance on the PDBbind v.2016 set and two independent sets. (A) PDBbind validation set. (B) PDBbind v.2016 core set. (C) CASF-2013. (D) Astex Diverse set.

For classification task, the model was evaluated on DUD-E, *Human*, *C. elegans* and KIBA datasets using 3-fold cross validation. Two traditional machine learning algorithms, SVM and random forest, were also used to compare on these datasets. As shown in Fig. 3, the proposed model has shown excellent performance on most evaluation metrics including AUC, accuracy, precision, recall, F1-score and specificity. Specifically, the single-task method, which indicates training by each one task, has achieved better performance than the multi-task method on more specific metrics such accuracy and precision. While the multi-task method has showed better performance on recall, indicating that it covered more protein-ligand interaction data space. It should be noted that each single-task model corresponds to only one dataset (i.e., it performs well on one dataset but performs poorly on the other datasets). Therefore, the single-task model is likely to perform poorly on coronavirus specific dataset due to the generalizability issue, even though it performed well on the benchmark dataset. In contrast, multi-task model was trained once and achieved excellent results on all these datasets. These results suggest that the multi-task model has a better applicability by leveraging heterogenous datasets.

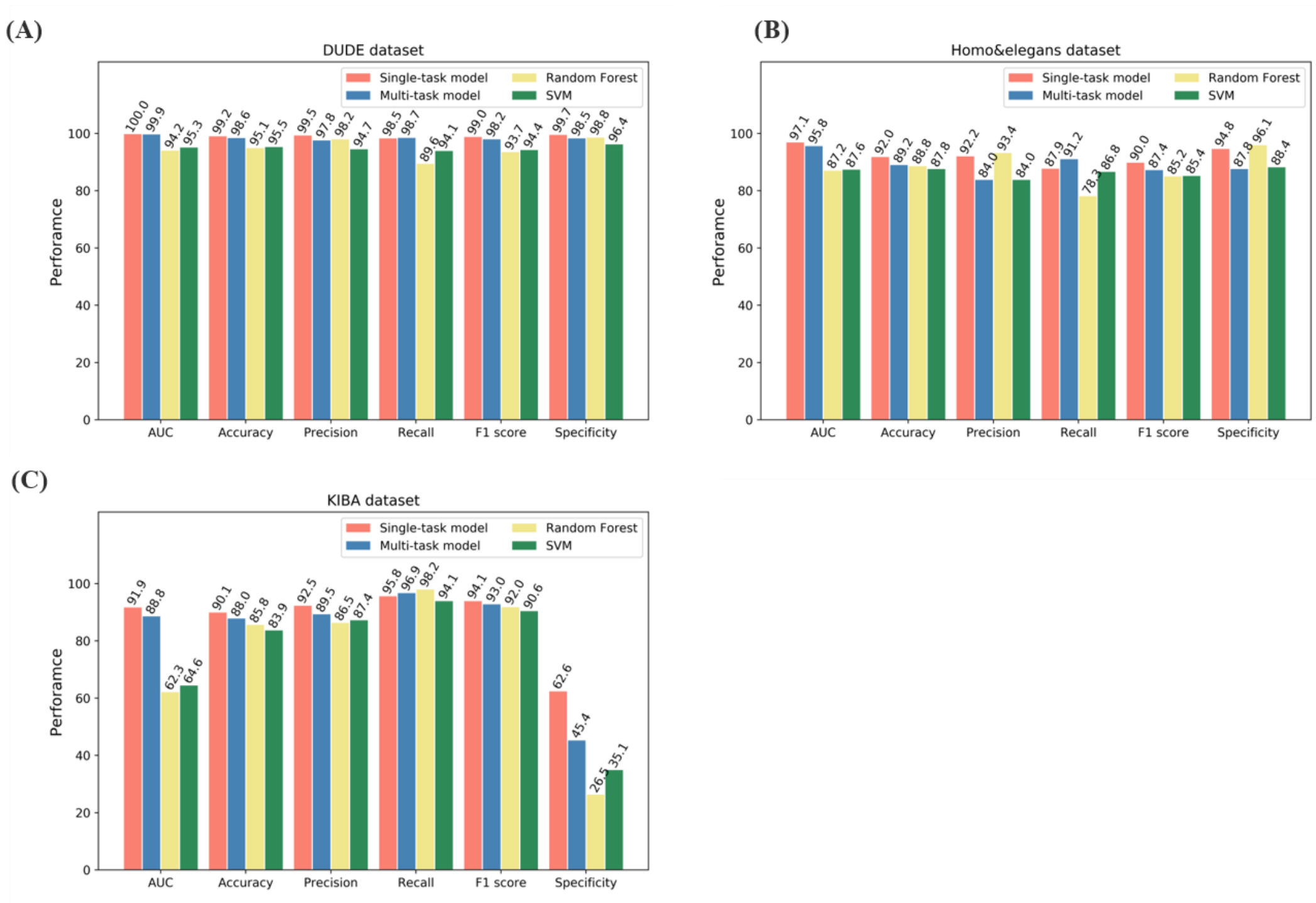

Fig. 3 Model performance on DUD-E, *Human*, *C. elegans* and KIBA datasets.

## Screening of SARS-CoV-2 inhibitors

Recently, pretraining model on a large-scale dataset before applying to a small dataset has emerged as a powerful paradigm for improving model generalizability. Inspired by this idea, we fine-tuned the pretrained model on a collected SARS-CoV-2 dataset and then selected potential SARS-CoV-2 inhibitors from a bioactive commercial library containing 10 thousand compounds. After excluding drugs with possible side effects, total 10 drugs with high binding affinity were listed in table 1. Among these, abacavir (sulfate), a powerful nucleoside analog reverse transcriptase inhibitor used to treat HIV, was predicted to have high binding affinity with multiple proteins of SARS-CoV-2 including RdRp and helicase. Darunavir, a protease inhibitor used to treat HIV, was used in a clinical trial against COVID-19 (ChiCTR2000029541). It should be noted that, both darunavir and darunavir (ethanolate) were not present in training and fine-tuning set. That is, they were "unseen" for the model. Our model predicted that darunavir can target to 3CLpro and PLpro with affinity $K_d$= 57.30 and 46.16 nM respectively. While darunavir (ethanolate) binds to 3CLpro and PLpro with affinity $K_d$= 44.51 and 35.86 nM, respectively. These results partially prove accuracy and generalizability of our model. In our predictions, almitrine mesylate, which was a respiratory stimulant that enhances respiration, was used in the treatment of chronic obstructive pulmonary disease. While roflumilast has anti-inflammatory effects and was used as an orally administered drug for the treatment of inflammatory conditions of the lungs such as chronic obstructive pulmonary disease. These two predicted drugs were associated with respiratory symptoms that are the main clinical symptoms of COVID-19. Interestingly, kesuting syrup and keqing capsule were used in a trial for treatment of mild and moderate COVID-19 (ChiCTR2000029991). It is uncertain whether these drugs only help to alleviate clinical symptoms or have direct effect on virus. Daclatasvir was used against Hepatitis C Virus, which stops HCV viral RNA replication and protein translation by directly inhibiting HCV protein NS5A. In this study, the predicted binding affinity between daclatasvir and RdRp was 15.03nM. Fiboflapon sodium, a high affinity 5-lipoxygenase-activating protein inhibitor used for the treatment of asthma, was predicted to have potential affinity to PLpro with $K_d$= 197.63nM.

Table 1. The potential inhibitors for SARS-CoV-2

| **Drug** | **CAS** | **Target** | **Predicted affinity(nM)** |
|---|---|---|---|
| Abacavir (sulfate) | 188062-50-2 | RdRp | 3.03 |
| | | helicase | 3.06 |
| Darunavir | 206361-99-1 | 3CLpro | 57.30 |
| | | PLpro | 46.16 |
| Darunavir (Ethanolate) | 635728-49-3 | 3CLpro | 44.51 |
| | | PLpro | 35.86 |
| Itraconazole | 84625-61-6 | PLpro | 127.98 |
| | | RdRp | 16.90 |
| Almitrine mesylate | 29608-49-9 | 3CLpro | 29.31 |
| Daclatasvir | 1009119-64-5 | RdRp | 15.03 |
| Daclatasvir (dihydrochloride) | 1009119-65-6 | RdRp | 19.87 |
| Metoprolol tartrate | 56392-17-7 | PLpro | 153.23 |
| Fiboflapon sodium | 1196070-26-4 | PLpro | 197.63 |
| Roflumilast | 162401-32-3 | 3CLpro | 248.89 |

Furthermore, the model was applied to screen lead compounds for 3CLpro inhibition from a diverse screening library containing 2 million compounds. This screening libraries, namely Enamine HTS Collection, encompasses versatile chemotypes developed within a couple of decades of chemical research at Enamine and its partner academic organizations. Compounds within this library frequently have unusual structures and unique properties. 3CLpro was taken as the screening target because of its wide application and the largest amount of corresponding data. Thus, the fine-tuned deep learning model has relatively higher accuracy for selecting 3CLpro inhibitors. After model prediction, we selected 10 potential lead compounds based on probability and diversity (listed in Table 2).

Table 2. The predicted lead compounds for SARS-CoV-2 3CLpro

| **ID** | **SMILES** | **Probability** |
|---|---|---|
| Z56899184 | O=[N+]([O-])c1cccc(SSc2cccc([N+](=O)[O-])c2)c1 | 0.969 |
| Z229622170 | N#Cc1cccc(CN2C(=O)C(=O)c3cccc(Br)c32)c1 | 0.947 |
| Z57728899 | COc1ccc(CN2C(=O)C(=O)c3cc(Br)ccc32)cc1 | 0.918 |
| Z90667629 | [Na+].[O-][n+]1ccccc1[S-] | 0.863 |
| Z1238998507 | N#Cc1cc([N+](=O)[O-])ccc1Oc1cncc(Cl)c1 | 0.782 |
| Z56833036 | O=C1c2ccccc2C(=O)c2c1cccc2S(=O)(=O)N1CCOCC1 | 0.724 |
| Z57013003 | COc1ccc(N2CCN(C(=O)c3cc(=O)[nH]c4ccccc34)CC2)cc1 | 0.711 |
| Z1245218850 | N#CCc1cccc(C(=O)Oc2cncc(Cl)c2)c1 | 0.697 |
| Z56785091 | c1nc(SSc2nc[nH]n2)n[nH]1 | 0.682 |
| Z1776036493 | Cc1cc(Br)cc2c1N(CCBr)C(=O)C2=O | 0.676 |

## Model interpretation

To explore how the model discerns protein-ligand interaction, we conduct a method to identify the key amino acids that are critical for binding. The listed potential inhibitors were regarded as positive samples with high prediction score. Then we masked sub-sequences of sample to get a “masked” prediction score, and then calculated the importance of these masked sub-sequences. As shown in Fig.4, the critical parts for binding in protein sequences are visualized using heat-map. The yellower in color the region, the more important it is. For 3CLpro, the important amino acids for binding mostly located at two main parts. It should be noted that different drugs result in different weights of these two regions. For example, roflumilast has higher weights in the first region, indicating the binding sites for roflumilast are close to the middle pocket of 3CLpro. As for abacavir (sulfate), both regions more or less affect the binding, especially the second region at 180-200th amino acids. For PLpro, the predicted binding sites locate at 100-120th amino acids.

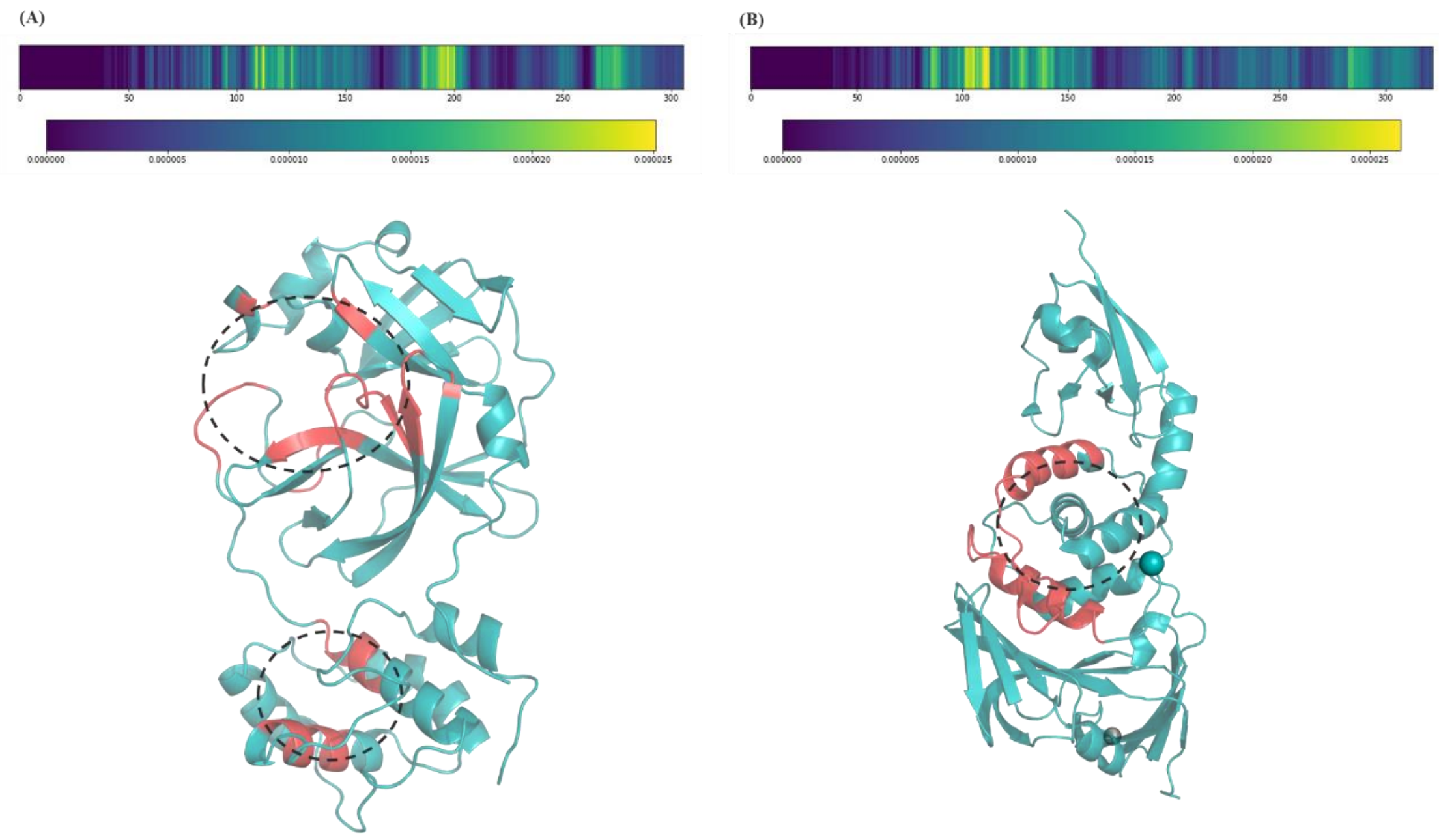

Fig. 4 The predicted binding sites. (A) 3CLpro (PDB ID: 6lu7); (B) PLpro (PDB ID: 6wx4). (The abscissa axis is the length of protein sequences.) Red indicates the predicted important sites and black dotted circles indicate possible pockets.

Moreover, we visualized the predicted binding sites in 3D structures to show possible pockets. The 3CLpro of SARS-CoV-2 (PDB: 6lu7) was used. As mentioned above, two regions of 3CLpro contribute mostly to the binding. As shown in Fig.4A, the region in the upper part (100-200th amino acids) was predicted as the main pocket due to the high weights in most predictions. SARS-CoV-2 PLpro has a catalytic triad composing of Cys114-His275-Asp289 and a conventional zinc-binding domain of 4 cysteine residues Cys192, Cys194, Cys227, Cys229[19]. The central "thumb" domain contributes the catalytic Cys112 to the active site. Similarly, our model predicted that 100-120th amino acids contributed mainly to the final binding of small molecules.

## Molecular Docking

Based on our prediction and known protein-ligand structural complexes, we performed molecular docking of predicted inhibitors to 3CLpro. Four inhibitors, Z56899184, Z57728899, Z57013003 and Z1245218850, were docked to 3CLpro using AutoDock Vina[20]. As shown in Fig. 5, these inhibitors all bind to amino acids that located in the active site of 3CLpro (the predicted upper binding region in Fig. 4A). Previous study showed that the catalytic residue Cys145 was critical for forming covalent bond between 3CLpro and inhibitors[21,22]. Although Z56899184 and Z57013003 were not covalent inhibitors for 3CLpro, their binding regions were closer to Cys145 which may probably result in higher binding affinity. But it should be noted that the binding affinity between protein and ligand cannot be simply considered based on docking scores (i.e., often expressed in kcal/mol).

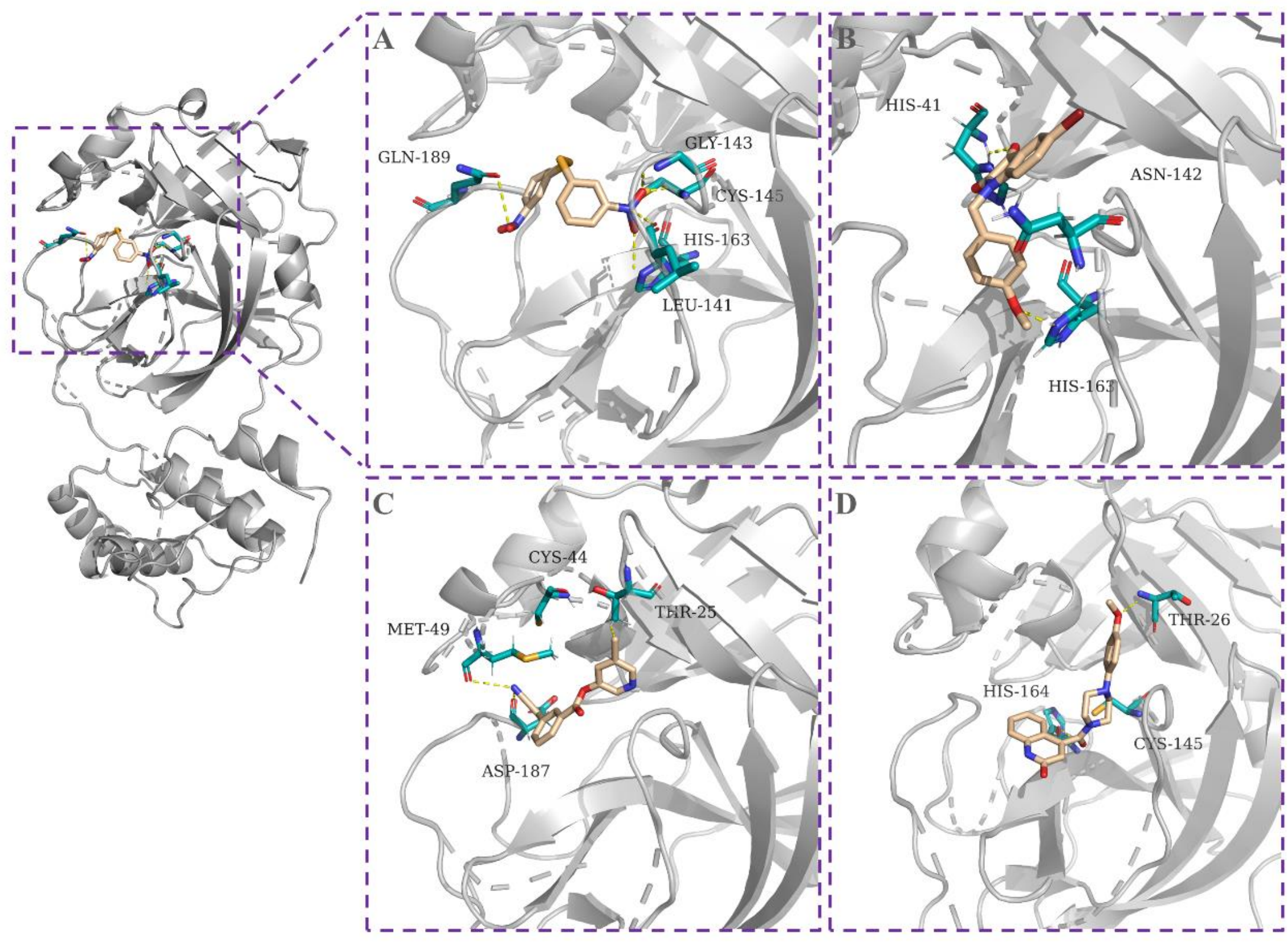

Fig. 5 Molecular docking of predicted inhibitors to SARS-CoV-2 3CLpro. (A) Z56899184, -6.5 kcal/mol; (B) Z57728899, -6.3 kcal/mol; (C) Z1245218850, -5.7 kcal/mol; (D) Z57013003, -7.4 kcal/mol. 3CLpro is represented as translucent gray cartoon. The involved amino acids are represented as cyan color

# Conclusion

In this work, a multi-task deep learning model was proposed to predict potential commercial drugs against SARS-CoV-2. There are two main points that should be restated. First, the model was pretrained on heterogenous protein–ligand interaction datasets before being applied to the SARS-CoV-2 inhibitor screening. This idea was inspired by the recent success of pretrained models in various fields, such as natural language processing. The core logic of this method was transfer learning, which could extract features from large-scale datasets and then transfer the learned knowledge to a small set. This method has a broad application in drug discovery, which is often hindered by the small data issue. Although the pretrained model had a wide distribution in the protein–ligand interaction data space, the distance between existing knowledge and future application is ambiguous. Therefore, which part of the pretraining set would cause issues such as negative effects was not understood well. Second, the exploration of the "black box" within deep learning models is very important, especially in the drug discovery area. The black box model may increase the risk of subsequent false leads, thus causing severe consequences. Our model could provide some insights into this. Overall, we are on a cheerful avenue for the application of pretrained deep learning models in drug discovery.

# Methods

## Data

A few benchmark protein-ligand interaction datasets were used for model pretraining, including DUD-E[23], *Human*, *C. elegans*[24], KIBA[25] and PDBbind[15]. After preprocessing, total 271,816 protein-ligand interactions were used in this study. For drug screening, several SARS-CoV-2 important proteins were used as targets such as RNA-dependent RNA polymerase (RdRp), 3-chymotrypsin-like protease, Papain-like protease and helicase. The sequences of these proteins were extracted from NCBI (NC_045512.2). The SARS-CoV/SARS-CoV-2 specific dataset was collected from various papers [21,26–29] and public datasets such as GHDDI and PubChem with filtration. After removing duplicates, a collection of 10 thousand approved and bioactive compounds was used for screening. Similarly, a diverse Enamine HTS collection containing 2 million compounds was also used.

## Model

Basically, the proposed model consists of three parts: protein feature extraction by word2vec and transformer, drug feature extraction by node2vec and interaction prediction by multi-task modules. The last dense layer is activated as output. The loss functions are defined as binary cross entropy (BCE) and mean squared error (MSE) for classification and regression, respectively. Fig. 1 displays the basic architecture of the model, more details are described below.

**Preprocessing.** Given the drug SMILES and protein sequence, we first construct graph and words respectively, and then get the dense vectors by embedding method. Let $L_{idx} = x_1, x_2, \ldots, x_{L-1}, x_L$ be a word sequence of each protein, where idx is the sequence index, L is the word sequence length and $x_i \in R^{d_1}$ is the d1-dimensional embedding of ith word that we split by overlapping 3-gram on original sequence. For drugs, let $G = (V, E)$ be a given drug SMILES, and $f: V \rightarrow R^{d_2}$ be the mapping function from $V_{idx} = \{v_1, v_2, \ldots, v_{l-1}, v_l\}$ to feature representation we aim to learn for protein-drug interaction prediction, where $idx$ is the sequence index, $l$ is the drug length and $v_i$ is the d2-dimensional embedding of $i$th node on original sequence[30].

**Pre-training.** With building of a large scale word segmented corpus of protein sequence, we then use each current word as an input to a log-linear classifier with continuous projection layer, and predict words within a certain range before and after the current word, proposed as skip-gram model[13], to obtain the final $d_1$-dimensional embedding words. Meanwhile, to identify the specific drug-network nodes, we consider node2vec[14] as the representation learning method, which design a flexible neighborhood sampling strategy to explore a random walk, such as $\{s, v_3, v_5, v_9, v_8, v_1, v_6, v_4\}$ of start node $s = v_2$ and walking length l = 8, thus we could consider the skip-gram architecture what we mentioned above as feature learning method.

**Multi-task module.** This module consists of two parts: shared layers and task-specific layers. The shared layers are designed to learn a joint representation for all tasks. Task-specific layers are used to learn the weights of specific blocks based on the joint representation. In this study, two related tasks are defined: classification and regression. The model was fine-tuned by virus-specific dataset to acquire robust results for coronavirus.

### Biological interpretation

A non-parametric method "occlusion" used in our previous study [11] was applied to explore which parts of the input protein sequences were critical to the task. Briefly, $s_i$ from test samples ($i = 0, 1, 2, ..., n-1$, here n is sample size of test set) was expressed as tuple (protein input_i, compound input_i). While maintaining compound input_i unchanged, we systematically masked the protein input_i in $s_i$ to track the changes of the output. Then the importance of each sub-sequence in the sequence to the prediction can be calculated.

## Acknowledgement

This work was supported by the Strategic Priority Research Program of Chinese Academy of Sciences (NO. XDB 38040200) and the Shenzhen Science and Technology Innovation Committee (JCYJ20180703145002040) and Shenzhen Science and Technology Program (Grant No. SGDX20201103095603009)